\documentclass[12pt,fleqn]{article}
\usepackage{graphics,epsfig}
\usepackage{palatino}

\begin{document}
\title{An information theoretic approach to the functional
classification of neurons}

\author{Elad Schneidman,$^{1,2}$ William Bialek,$^1$
and Michael J. Berry II$^2$\\
$^1$Department of Physics and $^2$Department of Molecular Biology\\
Princeton University, Princeton NJ 08544, USA\\
{\it \{elads,wbialek,berry\}@princeton.edu} } \maketitle

\begin{abstract}
A population of neurons typically exhibits a broad diversity of
responses to sensory inputs. The intuitive notion of functional
classification is that cells can be clustered so that most of the
diversity is captured in the identity of the clusters rather than
by individuals within clusters. We show how this intuition can be
made precise using information theory, without any need to
introduce a metric on the space of stimuli or responses.  Applied
to the retinal ganglion cells of the salamander, this approach
recovers classical results, but also provides clear evidence for
subclasses beyond those identified previously. Further, we find
that each of the ganglion cells is functionally unique, and that
even within the same subclass only a few spikes are needed to
reliably distinguish between cells.
\end{abstract}

\section{Introduction}

Neurons exhibit an enormous variety of shapes and molecular
compositions. Already in his classical work, Cajal \cite{Cajal-11}
recognized that the shapes of cells can be classified, and he
identified many of the cell types that we recognize today.   Such
classification is fundamentally important, because it implies that
instead of having to describe $\sim$10$^{12}$ individual neurons,
a mature neuroscience might need to deal only with a few thousand
different classes of nominally identical neurons.  There are three
broad methods of classification: morphological, molecular, and
functional.  Morphological and molecular classification are
appealing because they deal with a relatively fixed property, but
ultimately  the functional properties of neurons  are the most
important, and neurons that share the same morphology or molecular
markers need not embody the same function. With attention to
arbitrary detail, every neuron will be individual, while  a
coarser view might overlook an important distinction; a
quantitative formulation of the classification problem is
essential.

The vertebrate retina is an attractive example: its anatomy is
well studied and highly ordered, containing repeated
micro-circuits that look out at different angles in visual space
\cite{Cajal-11,Dowling-87}; its overall function (vision) is
clear, giving the experimenter better intuition about relevant
stimuli; and responses of many of its output neurons, ganglion
cells, can be recorded simultaneously using a multi--electrode
array, allowing greater control of experimental variables than
possible with serial recordings \cite{Meister-Pine-Baylor-94}.
Here we exploit this favorable experimental situation to highlight
the mathematical questions that must lie behind any attempt at
classification.

Functional classification of retinal ganglion cells typically has
consisted of finding qualitatively different responses to simple
stimuli.  Classes are defined by whether ganglion cells fire
spikes at the onset or offset of a step of light or both (ON, OFF,
ON/OFF cells in frog \cite{Hartline-37}) or whether they fire once
or twice per cycle of a drifting grating (X, Y cells in cat
\cite{Hochstein-Shapley-76}). Further elaborations exist.  In the
frog, the literature reports 1 class of ON-type ganglion cell and
4 or 5 classes of OFF-type \cite{Grosser-Cornehls-76}.  The
salamander has been reported to have only 3 of these OFF-type
ganglion cells \cite{Grosser-Cornehls-Himstedt-73}.  The classes
have been distinguished using stimuli such as diffuse flashes of
light, moving bars, and moving spots.  The results are similar to
earlier work using more exotic stimuli \cite{Lettvin-59}.  In some
cases, there is very close agreement between anatomical and
functional classes, such as the ($\alpha$,$\beta$) and (Y,X) cells
in the cat.  However, the link between anatomy and function is not
always so clear.

Here we show how information theory allows us to define the
problem of classification without any a priori assumptions
regarding which features of visual stimulus or neural response are
most significant, and without imposing  a metric on these
variables. All notions of similarity emerge from the joint
statistics of neurons in a population as they respond to common
stimuli.  To the extent that we identify the function of retinal
ganglion cells as providing the brain with information about the
visual world, then our approach finds exactly the classification
which captures this functionality in a maximally efficient manner.
Applied to experiments on the tiger salamander retina, this method
identifies the major types of  ganglion cells in agreement with
traditional methods, but on a finer level we find clear structure
within a group of 19 fast OFF cells that suggests at least 5
 functional subclasses.  More profoundly, even cells within a subclass
 are very different from one another, so that on average the
ganglion cell responses to the simplified visual stimuli we have
used provide $\sim$6 bits/sec of information about cell identity
within our cell population. This is sufficient to identify
uniquely each neuron in an ``elementary patch'' of the retina
within one second, and a typical pair of cells can be
distinguished reliably by observing an average of just two or
three spikes.

\section{Theory}

Suppose that we could give a complete characterization, for each
neuron ${\rm i} = 1, 2, \cdots , N$ in a population, of the
probability $P(r | {\bf \vec s}, {\rm i})$ that a stimulus ${\bf
\vec s}$ will generate the response $r$.  Traditional approaches
to functional  classification introduce (implicitly or explicitly)
a parametric representation for the distributions $P(r | {\bf \vec
s}, {\rm i})$ and then search for clusters in parameter space. For
visual neurons we might assume that responses are determined by
the projection of the stimulus movie ${\bf \vec s}$ onto a single
template or receptive field, $P(r | {\bf \vec s}, {\rm i}) = F(r;
{\bf \vec f}_{\rm i} {\bf \cdot} {\bf \vec s}) $; classifying
neurons then amounts to clustering the receptive fields ${\bf \vec
f}_{\rm i}$.  But it is not possible to cluster without specifying
what it means for these vectors to be similar; in this case, since
the vectors come from the space of stimuli, we need a metric or
distortion measure on the stimuli themselves.  It seems strange
that classifying the responses of visual neurons requires us to
say in advance what it means for images or movies to be
similar.\footnote{If all cells are selective for  a small number
of commensurate features, then the set of vectors ${\bf \vec
f}_{\rm i}$ must lie on a low dimensional manifold, and we can use
this selectivity to guide the clustering.  But we still face the
problem of defining similarity:  even if all the receptive fields
in the retina can be summarized meaningfully by the diameters of
the center and surround (for example), why should we believe that
Euclidean distance in this two dimensional space is a sensible
metric?}

Information theory suggests a formulation that does not require us
to measure similarity among either   stimuli or  responses.
Imagine that we  present a stimulus ${\bf \vec s}$ and record the
response $r$ from a single neuron in the population, but we don't
know which one.   This response tells us something about the
identity of the cell, and on average this can be quantified as the
mutual information between responses and identity (conditional on
the stimulus),
\begin{equation}
I(r;{\rm i}|{\bf \vec s}) = {1\over N} \sum_{{\rm i}=1}^N \sum_r
P(r|{\bf \vec s}, {\rm i}) \log_2\left[ {{P(r|{\bf \vec s}, {\rm
i})}\over{P(r|{\bf \vec s})}} \right]\, {\rm bits},
\end{equation}
\noindent where $P(r|{\bf \vec s}) = (1/N) \sum_{{\rm i}=1}^N
P(r|{\bf \vec s}, {\rm i})$. The mutual information  $I(r;{\rm
i}|{\bf \vec s})$ measures the extent to which different cells in
the population produce {\em reliably} distinguishable responses to
the same stimulus; from Shannon's classical arguments
\cite{Shannon-Weaver-49} this is the unique measure of these
correlations which is consistent with simple and plausible
constraints. It is  natural to ask this question on average in an
ensemble of stimuli $P({\bf \vec s})$ (ideally the natural
ensemble),
\begin{equation}
\langle I(r;{\rm i} |{\bf \vec s}) \rangle_{\bf \vec s} = {1\over
N} \sum_{{\rm i}=1}^N \int [d{\bf\vec s}] P({\bf \vec s}) P(r|{\bf
\vec s}, {\rm i}) \log_2\left[ {{P(r|{\bf \vec s}, {\rm
i})}\over{P(r|{\bf \vec s})}} \right] ;
\end{equation}
$\langle I(r;{\rm i} |{\bf \vec s}) \rangle_{\bf \vec s}$ is
invariant under all invertible transformations of $r$ or $\bf \vec
s$.

Because information is mutual, we also can think of $\langle
I(r;{\rm i} |{\bf \vec s}) \rangle_{\bf \vec s}$ as the
information that cellular identity provides about the responses we
will record.  But now it is clear what we mean by classifying the
cells:  If there are clear classes, then we can predict the
responses to a stimulus just by knowing the class to which a
neuron belongs rather than knowing its unique identity.  Thus we
should be able to find a mapping ${\rm i} \rightarrow C$ of cells
into classes $C = 1, 2, \cdots , K$ such that $\langle I(r;C |{\bf
\vec s}) \rangle_{\bf \vec s}$ is almost as large as $\langle
I(r;{\rm i} |{\bf \vec s}) \rangle_{\bf \vec s}$, despite the fact
that the number of classes $K$ is much less than the number of
cells $N$.

Optimal classifications are those which use the $K$ different
class labels to capture as much information as possible about the
stimulus-response relation, maximizing $\langle I(r;C |{\bf \vec
s}) \rangle_{\bf \vec s}$ at fixed $K$.  More generally we can
consider soft classifications, described by probabilities
$P(C|{\rm i})$ of assigning each cell to a class, in which case we
would like to capture as much information as possible about the
stimulus-response relation while constraining the amount of
information that class labels provide directly about identity,
$I(C;{\rm i})$.  In this case our optimization problem becomes,
with $T$ as a Lagrange multiplier,
\begin{equation}
\max_{P(C|{\rm i})} \left[ \langle I(r;C |{\bf \vec s})
\rangle_{\bf \vec s} - T I(C;{\rm i}) \right].
\end{equation}
This is a generalization of the information bottleneck problem
\cite{Tishby-Pereira-Bialek-99}.

Here we confine ourselves to hard classifications, and use a
greedy agglomerative algorithm \cite{Slonim-Tishby-00} which
starts with $K=N$ and makes mergers which at every step provide
the smallest reduction in $I(r;C |{\bf \vec s})$.  This
information loss on merging cells (or clusters) ${\rm i}$ and
${\rm j}$ is given by
\begin{equation}
D({\rm i} , {\rm j}) \equiv \Delta I_{\rm ij}(r;C |{\bf \vec s}) =
{\large\langle} D_{JS}[P(r|{\bf \vec s}, {\rm i})  ||P(r|{\bf \vec
s}, {\rm j}) ] {\large\rangle}_{\bf\vec s} , \label{eq:DeltaI}
\end{equation}
where $D_{JS}$ is the Jensen--Shannon divergence \cite{Lin-91}
between the two distributions, or equivalently the information
that one sample provides about its source distribution in the case
of just these two alternatives. The matrix of ``distances''
$\Delta I_{\rm ij}$ characterizes the similarities among neurons
in pairwise fashion.

Finally, if cells belong to clear classes, then we ought to be
able to replace each cell by a typical or average member of the
class without sacrificing function.  In this case function is
quantified by asking how much information cells provide about the
visual scene.  There is a strict complementarity of the
information measures:  information that the stimulus/response
relation provides about the identity of the cell is exactly
information about the visual scene which will be lost if we don't
know the identity of the cells
\cite{Schneidman-Brenner-Tishby-dRvS-Bialek-01}.  Our information
theoretic approach to classification of neurons thus produces
classes such that replacing cells with average class members
provides the smallest loss of information about the sensory
inputs.

\section{The responses of retinal ganglion cells to identical stimuli}

We recorded simultaneously from 21 retinal ganglion cells from the
salamander using a multi-electrode array.\footnote{The retina is
isolated from the eye of the larval tiger salamander ({\em
Ambystoma tigrinum tigrinum}) and perfused in Ringer's medium.
Action potentials were measured extracellularly using a
multi-electrode array \cite{Meister-Pine-Baylor-94}, while light
was projected from a computer monitor onto the photoreceptor
layer. Because erroneously sorted spikes would strongly effect our
results, we were very conservative in our identification of
cleanly isolated cells.} The visual stimulus consisted of 100
repeats of a 20 s segment of spatially uniform flicker (see
fig.\,1a), in which light intensity values were randomly selected
every 30 ms from a Gaussian distribution having a mean of 4
mW/mm$^2$ and an RMS contrast of 18\%. Thus, the photoreceptors
were presented with exactly the same visual stimulus, and the
movie is many correlation times in duration, so we can replace
averages over stimuli by averages over time (ergodicity).  A 3 s
sample of the ganglion cell's responses to the visual stimulus is
shown in Fig.\,1b. There are times when many of the cells fire
together, while at other times only a subset of these cells is
active. Importantly, the same neuron may be part of different
active groups at different times. On a finer time scale than shown
here, the latency of the responses of the single neurons and their
spiking patterns differ across time. To analyze the responses of
the different neurons, we discretize the spike trains  into time
bins of size $\Delta t$. We examine the response  in windows of
time having length $T$, so that an individual neural response $r$
becomes a binary `word' $W$ with $T/\Delta t$
`letters'.\footnote{As any fixed choice of $T$ and $\Delta t$ is
arbitrary, we explore a range of these parameters.}

\clearpage

\begin{figure}[hbt]
%   \vspace{-1.25cm}
    \vspace{-1.25cm}
    \hspace{0.35cm}
    \includegraphics[width=12cm]{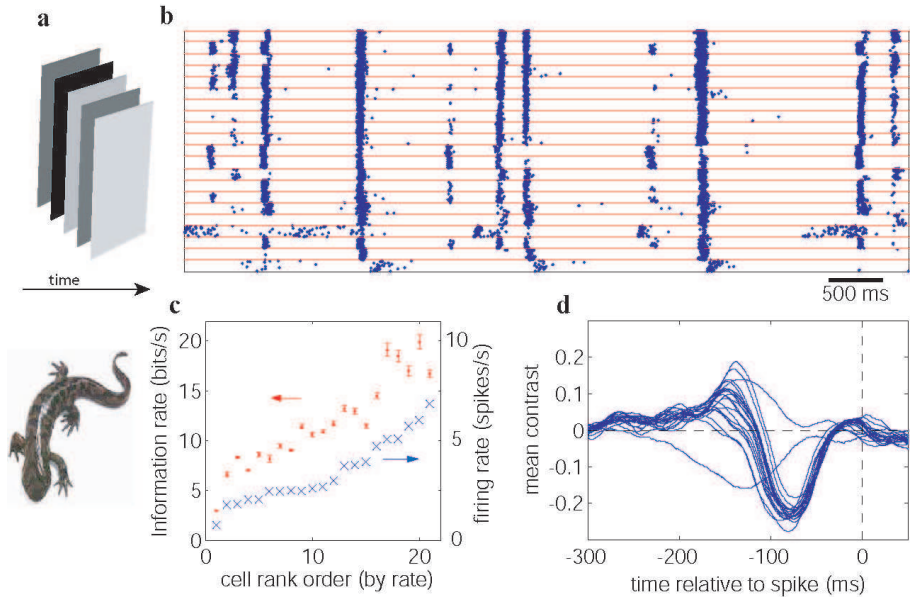}\\
    \vspace{-0.2cm}
\caption{ \small{ {\bf Responses of salamander
 ganglion cells to modulated uniform field intensity}. {\bf a}: The retina
 is presented with a series of uniform intensity ``images". The intensity
 modulation is Gaussian white noise distributed. {\bf b}: A 3 sec segment
 of the (concurrent) responses of 21 ganglion cells to repeated
 presentation of the stimulus. The rasters are ordered from bottom to top
 according to the average firing rate of the neurons (over the whole
 movie). {\bf c}: Firing rate and Information rates of the different cells
 as a function of their rank, ordered by their firing rate. {\bf d}: The
 average stimulus pattern preceding a spike for each of the different
 cells. Traditionally, these would be classified as 1 ON cell, 1 slow-OFF
 cell and 19 fast-OFF cells.}}
  \end{figure}

Since the cells in Fig.\,1b  are ordered according to their
average firing rate, it is clear that there  is no `simple'
grouping of the cells' responses with respect to this response
parameter; firing rates range continuously from 1 to 7 spikes per
second (Fig.\,1c). Similarly,  the rate of information (estimated
according to \cite{Strong-Koberle-dRvS-Bialek-98}) that the cells
encode about the same stimulus also ranges continuously from 3 to
20 bits/s. We estimate the average stimulus pattern preceding a
spike for each of the cells,  the spike triggered average (STA),
shown in Fig.\,1d. According to traditional classification based
on the STA, one of the cells is an ON cell, one is a slow OFF
cells and 19 belong to the fast OFF class
\cite{Keat-Reinagel-Reid-Meister-01}. While it may be possible to
separate the 19 waveforms of the fast OFF cells into subgroups,
this requires assumptions about what stimulus features are
important.  Furthermore, there is no clear standard for ending
such subclassification.

\section{Clustering of the ganglion cells responses into functional types}

To classify these ganglion cells  we solved the information
theoretic optimization problem described above.  Figure 2a shows
the pairwise distances $D({\rm i}, {\rm j})$ among the 21 cells,
ordered by their average firing rates; again,  firing rate alone
does not cluster the cells.  The result of the greedy clustering
of the cells is shown by a binary dendrogram  in Fig.\,2b.

\begin{figure}[hbt]
%   \vspace{-0.75cm}
    \vspace{-0.35cm}
   \hspace{0.2cm}
   \includegraphics[width=12cm]{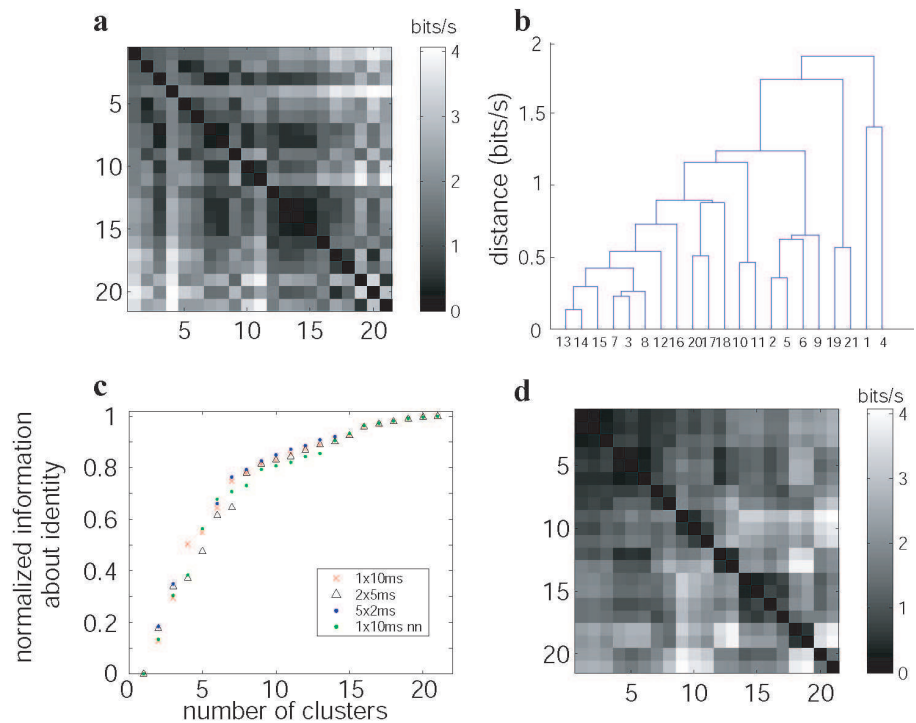}\\
%    \vspace{-0.75cm}
   \vspace{-0.5cm}
\caption{\small{ {\bf Clustering ganglion cell responses}. {\bf
a}: Average distances between the cells responses; cells are
ordered by their average firing rate. {\bf b}: Dendrogram of cell
clustering. Cell names correspond to their firing rate rank. The
height of a merge reflects the distance between merged elements.
{\bf c}: The information that the cells' responses convey about
the clusters in every stage of the clustering in (b), normalized
to the total information that the responses convey about cell
identity. Using different response segment parameters or
clustering method (e.g., nearest neighbor) result in very similar
behavior. {\bf d}: reordering of the distance matrix in (a)
according to the tree structure given in (b). }}
 \end{figure}

\clearpage

The greedy agglomerative approximation \cite{Slonim-Tishby-00}
starts from every cell as a single cluster.  We iteratively merge
the clusters $c_i$ and $c_j$ which have the minimal value of
$D(c_i,c_j)$ and display this distance or information loss as the
height of the merger in Fig.\,2b.  We pool their spike trains
together as the responses of the new cell class. We now
re-estimate the distances between clusters and repeat the
procedure, until we get a single cluster that contains all cells.
Fig.\,2c shows the compression in information achieved by each of
the mergers: for each number of clusters, we plot the mutual
information between the clusters and the responses, $\langle I(r;C
|{\bf \vec s}) \rangle_{\bf \vec s}$, normalized by the
information that the response conveys about the full set of cells,
$\langle I(r;{\rm i} |{\bf \vec s}) \rangle_{\bf \vec s}$. The
clustering structure and the information curve in Fig.\,2c are
robust (up to one cell difference in the final dendrogram) to
changes in the word size and bin size used; we even obtain the
same results with   a nearest neighbor clustering based on
$D(i,j)$. This suggests that the top 7 mergers in Fig.\,2b (which
correspond to the bottom 7 points in panel c) are of significantly
different subgroups.  Two of these mergers, which correspond to
the rightmost branches of the dendrogram, separate out the ON and
slow OFF cells.  The remaining 5 clusters are subclasses of fast
OFF cells. However, Fig.\,2d which shows the dissimilarity matrix
from panel a, reordered by the result of the clustering,
demonstrates that while there is clear structure within the cell
population, the subclasses there are not sharply distinct.

 \subsection*{How many types are there?}

While one might be happy with classifying the  fast OFF cells into
5 subclasses, we further asked whether the cells within a subclass
are reliably distinguishable from one another; that is, are the
bottom mergers in Fig.\,2b-c significant? We therefore randomly
split each of the 21 cells into 2 halves (of 50 repeats each), or
`siblings', and re-clustered. Figure 3a shows the resulting
dendrogram of this clustering, indicating that the cells are
reliably distinguishable from one another: The nearest neighbor of
each new half--cell is its own sibling, and (almost) all of the
first layer  mergers are of the corresponding siblings (the only
mismatch is of a sibling merging with a neighboring full cell and
then with the other sibling). Figure\,3b shows the very different
cumulative probability distributions of pairwise distances among
the parent cells and that of the distances between siblings.

\clearpage

  \begin{figure}[hbt]
%   \vspace{-0.5cm}
   \vspace{-0.5cm}
%  \hspace{0.5cm}
%   \includegraphics[scale=0.275, bb= 20 20 575 575]{F3_5c.jpg}\\
   \includegraphics[width=12cm]{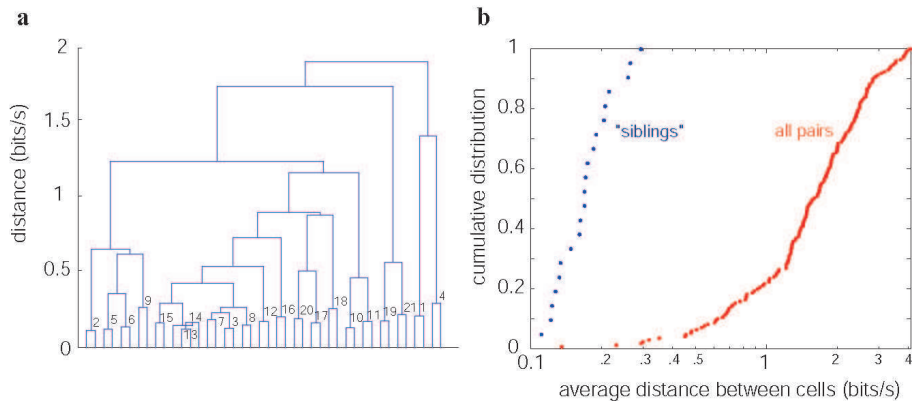}\\
   \vspace{-0.2cm}
\caption{\small {\bf Every cell is different than the
others}. {\bf a}: Clustering of cell responses after randomly
splitting every cell into 2 ``siblings". The nearest neighbor of
each of the new cells is his sibling and (except for one case) so
is the first merge. From the second level upwards, the tree is
identical to Fig.\,2b (up to symmetry of tree plotting). {\bf b}:
Cumulative distribution of pairwise distances between cells. The
distances between siblings are easily discriminated from the
continuous distribution of values of all the (real) cells.}
 \end{figure}

\subsection*{How significant are the differences between the cells?}

It might be that cells are distinguishable,  but only after
observing their responses for very long times. Since 1 bit is
needed to reliably distinguish between a pair of cells, Fig.\,3b
shows that more than 90\% of the pairs are reliably
distinguishable within 2 seconds or less. This result is
especially striking given the low mean spike rate of these cells;
clearly, at times where none of the cells is spiking, it is
impossible to distinguish between them. To place the information
about identity on an absolute scale, we compare it to the entropy
of the responses at each time, using 10 ms segments of the
responses at each time during the stimulus (Fig.\,4a). Most of the
points lie close to the origin, but many of them reflect discrete
times when the responses of the neurons are very different and
hence highly informative about cell identity: under the conditions
of our experiment, roughly 30\% of the response variability among
cells is informative about their identity.\footnote{Since the
cells receive the same stimulus and often possess shared
circuitry, an efficiency as high as 100\% is very unlikely.} On
average observing a single neural response gives about 6 bits/s
about the identity of the cells within this population. We also
compute the average number of spikes per cell which we need to
observe to distinguish reliably between cells ${\rm i}$ and ${\rm
j}$,
\begin{equation}
n_d(i,j)=\frac{\frac{1}{2} (\bar{r_i}+\bar{r_j})}{D(i,j)}.
\end{equation}

\noindent where $\bar{r_i}$ is the average spike rate of cell $i$
in the experiment. Figure\,4b shows the cumulative probability
distribution of the values of $n_d$. Evidently, more than 80\% of
the pairs are reliably distinguishable after observing, on
average, only 3 spikes from one of the neurons. Since ganglion
cells fire in bursts, this suggest that most cells are reliably
distinguishable based on a single firing `event'! We also show
that for the 11 most similar cells (those in the left subtree in
Fig.\,2b) only a few more spikes, or one extra firing event, are
required to reliably distinguish them.

\begin{figure}[hbt]
%   \vspace{-0.5cm}
%   \hspace{0.5cm}
%   \includegraphics[scale=0.275, bb= 20 20 575 575]{F4_2c.jpg}\\
   \includegraphics[width=12cm]{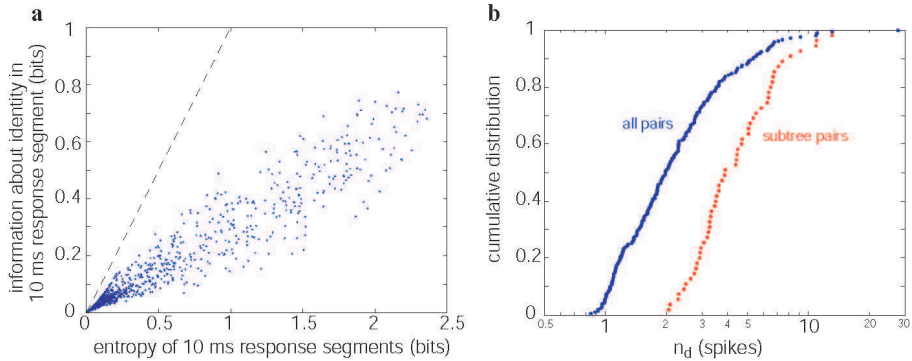}\\
   \vspace{-0.2cm}
\caption{\small {\bf Diversity is high} {\bf a}: The
average information that a response segment conveys about the
identity of the cell as a function of the entropy of the
responses. Every point stands for a time point along the stimulus.
Results shown are for 2-letter words of 5 ms bins; similar
behavior is observed for different word sizes and bins {\bf b}:
Cumulative distribution of the average number of spikes that are
needed to distinguish between pair of cells.}
 \end{figure}

\section{Discussion}

We have identified a diversity of functional types of retinal
ganglion cells by clustering them  to preserve information about
their identity. Beyond the easy classification of the major types
of salamander ganglion cells -- fast OFF, slow OFF, and ON -- in
agreement with traditional methods, we have found clear structure
within the fast OFF cells that suggests at least 5 more functional
classes. Furthermore, we found evidence that each cell is
functionally unique. Even under this relatively simple stimulus,
the analysis revealed that the cell responses convey $\sim$6
bits/s of information about cell identity within this population
of 21 cells. Ganglion cells in the salamander interact with each
other and collect information from a $\sim$250 $\mu$m radius;
given the density of ganglion cells, the observed rate implies
that a single ganglion cell can be discriminated from all the
cells in this ``elementary patch'' within 1 s. This is a
surprising degree of diversity, given that 19 cells in our sample
would be traditionally viewed as nominally the same.

One might wonder if our choice of uniform flicker  limits the
results of our classification. However, we found that this
stimulus was rich enough to distinguish every ganglion cell in our
data set. It is likely that stimuli with spatial structure would
reveal further differences. Using a larger collection of cells
will enable us to explore the possibility that there is a
continuum of unique functional units in the retina.

How might the brain make use of this diversity?   Several
alternatives are conceivable.  By comparing the spiking of closely
related cells, it might be possible to achieve much finer
discrimination among stimuli that tend to activate both cells.
Diversity also can  improve the robustness of retinal signalling:
as the retina is constantly setting its adaptive state in response
to statistics of the environment that it cannot estimate without
some noise, maintaining functional diversity can guard against
adaptation that overshoots its optimum. Finally, great functional
diversity opens up additional possibilities for learning
strategies, in which downstream neurons select the most useful of
its inputs rather than merely summing over identical inputs to
reduce their noise. The example of the invertebrate retina
demonstrates that nature can construct neural circuits with almost
crystalline reproducibility from synapse to synapse.  This
suggests that the extreme diversity found here in the vertebrate
retina may not be the result of some inevitable sloppiness of
neural development but rather as evolutionary selection of a
different strategy for representing the visual world.

\small{

}

\end{document}